\documentclass[aps,prc,twocolumn,groupedaddress,a4paper]{revtex4-1}

\usepackage{amsmath}
\usepackage{amssymb}
\usepackage{relsize}
\usepackage{graphicx}
\usepackage{textcomp}

\newcommand {\nc} {\newcommand}
\nc {\beq} {\begin{eqnarray}}
\nc {\eeq} {\end{eqnarray}}
\nc {\eeqn} [1] {\label{#1} \end{eqnarray}}
\nc {\eol} {\nonumber \\}
\nc {\ve} [1] {\mbox{\boldmath $#1$}}
\nc {\la} {\mbox{$\langle$}}
\nc {\ra} {\mbox{$\rangle$}}

\begin{document}

\title{Nonlocal nucleon-nucleus interactions in $(d,p)$ reactions:
Role of the deuteron $D$-state}

\author{G. W. Bailey}
\affiliation{Department of Physics, Faculty of Engineering and Physical
Sciences, University of Surrey Guildford, Surrey GU2 7XH, United Kingdom}
\author{N. K. Timofeyuk}
\affiliation{Department of Physics, Faculty of Engineering and Physical
Sciences, University of Surrey Guildford, Surrey GU2 7XH, United Kingdom}
\author{J. A. Tostevin}
\affiliation{Department of Physics, Faculty of Engineering and Physical
Sciences, University of Surrey Guildford, Surrey GU2 7XH, United Kingdom}

\date{\today}

\begin{abstract}
Theoretical models of the $(d,p)$ reaction are exploited for both nuclear
astrophysics and spectroscopic studies in nuclear physics. Usually, these
reaction models use local optical model potentials to describe the nucleon-
and deuteron-target interactions. Within such a framework the importance
of the deuteron $D$-state in low-energy reactions is normally associated
with spin observables and tensor polarization effects - with very minimal
influence on differential cross sections. In contrast, recent work that
includes the inherent nonlocality of the nucleon optical model potentials
in the Johnson-Tandy adiabatic-model description of the $(d,p)$ transition
amplitude, which accounts for deuteron break-up effects, shows sensitivity
of the reaction to the large n-p relative momentum content of the deuteron
wave function. The dominance of the deuteron $D$-state component at such
high momenta leads to significant sensitivity of calculated $(d,p)$ cross
sections and deduced spectroscopic factors to the choice of deuteron wave
function [Phys. Rev. Lett. {\bf 117}, 162502 (2016)]. We present details of
the Johnson-Tandy adiabatic model of the $(d,p)$ transfer reaction generalized
to include the deuteron $D$-state in the presence of nonlocal nucleon-target
interactions. We present exact calculations in this model and compare these
to approximate (leading-order) solutions. The latter, approximate solutions
can be interpreted in terms of local optical potentials, but evaluated at a
shifted value of the energy in the nucleon-target system. This energy shift
is increased when including the $D$-state contribution. We also study the
expected dependence of the $D$-state effects on the separation energy and
orbital angular momentum of the transferred nucleon. Their influence on the
spectroscopic information extracted from $(d,p)$ reactions is quantified
for a particular case of astrophysical significance.
\end{abstract}

\maketitle

\section{Introduction\label{intro}}
A primary interest in $(d,p)$ reactions, especially those studied at modern
radioactive beam facilities, is their ability to reveal single-particle spectra
of rare isotopes and to determine the angular momentum content and spectroscopic
strength of single-particle states near their Fermi-surfaces. This information
is crucial also in nuclear astrophysics applications. These angular momenta and
associated spectroscopic strengths are deduced from comparisons of the measured
cross sections with theoretical predictions. Differences between theoretical
models of the reaction thus impact the interpretation of experimental data and
studies of the sensitivities of calculations to model assumptions are vital.

Necessary inputs to direct reaction models of the $A(d,p)B$ transfer process,
such as the distorted-waves Born approximation (DWBA) \cite{Austern} and adiabatic
distorted-waves approximation (ADWA) \cite{JT} methods, are complex effective
interactions (optical potentials) between the reactants in the entrance ($d-A$)
and exit ($p-B$) channels. Feshbach theory clarifies that these interactions should
be both complex and nonlocal, arising from the many-body nature of the nuclei $A$
and $B$ and the effects of inelastic channel couplings upon the elastic channel
wave functions \cite{Fesh}. Within the DWBA, taking account of nonlocalities of
the Perey-Buck type \cite{PBuck} results in a multiplication of the entrance and
exit channel distorted waves by a {\em Perey factor} \cite{Per63}. However, the
DWBA neglects important contributions due to transfer from the deuteron breakup
continuum that require a consideration of the nucleon-target degrees of freedom
in the entrance channel. These deuteron breakup effects are treated efficiently
in the ADWA method \cite{JT}, that develops the $d-A$ effective potential from
those of the $n-A$ and $p-A$ systems. Until very recently, these $n-A$ and $p-A$
optical potentials used to describe the $n+p+A$ entrance channel were assumed to be
local. In the ADWA, nonlocality in the proton (exit) channel can be included in
the same manner as in the DWBA. This proton channel nonlocality has been treated
exactly in recent calculations \cite{Tit14,Ros15}. However, constructing the $d-A$
effective potentials when including nonlocal nucleon-target ($N-A$) potentials
required additional formal developments, as presented only relatively recently
\cite{Tim13a,Tim13b,Tit16,Ros16}. In addition, earlier work that included $N-A$
nonlocalities using Faddeev framework three-body calculations showed an improved
description of $(d,p)$ reaction cross sections on a range of closed-shell targets
\cite{Del09}.

It was shown in Refs. \cite{Tim13a,Tim13b} that including nonlocal $p-A$ and
$n-A$ potentials in the adiabatic model of the $A(d,p)B$ reaction generates a
nonlocal adiabatic $d-A$ potential. This model could be further reduced, to a
local one, in a similar way to that originally introduced by Perey and Buck
\cite{PBuck}. This revised local adiabatic potential $U_{dA}^{\rm loc}$ is
different to that which is usually constructed in the ADWA method, that uses
energy-dependent phenomenological local nucleon optical potentials and then
evaluates these at half the energy of the incident deuteron, $E_d$. Instead,
Refs. \cite{Tim13a,Tim13b} show, for $Z=N$ targets, that $U_{dA}^{\rm loc}$
should be constructed from nucleon optical potentials evaluated at a shifted
energy $E=E_d/2+\Delta E$. The required shift, $\Delta E$, is related to the
value of $\la T_{np}\ra_V$, a measure of the $n-p$ relative kinetic energy
$T_{np}$ in the deuteron ground-state $\phi_0$ inside the range of the $n-p$
interaction, $V_{np}$, that binds the deuteron. Specifically, this value is
\beq
\la T_{np}\ra_V = \frac{\langle \phi_0|V_{np}T_{np}|\phi_0\rangle}{\langle\phi_0|
V_{np}|\phi_0\rangle} \equiv\langle\phi_1|T_{np}|\phi_0\rangle ,
\eeqn{Tnp}
where we have defined
\beq
|\phi_1 \rangle = V_{np}|\phi_0 \ra /\langle\phi_0 |V_{np} | \phi_0 \ra.
\eeqn{phi1}
Given the short-range nature of the nucleon-nucleon (NN) interaction and $\phi_1$,
a major contribution to $\la T_{np}\ra_V $ arises from high $n-p$ relative momenta.

In Ref. \cite{Tim13b}, values of $\la T_{np}\ra_V $ and $\Delta E $ were obtained
assuming the $S$-state wave function of the purely attractive, phenomenological central
NN interaction of Hulth\'en \cite{Hul}, whereas realistic deuteron wave functions
have a modest $D$-state component with probability $P_D \approx 4-7\%$. Though modest,
this D-state component can dominate the wave function at high $n-p$ momenta with
important implications for calculations of $U_{dA}^{\rm loc}$ and $(d,p)$ cross sections.
The intrinsic nonlocality of optical potentials thus presents a distinct and novel
source of $D$-state and $n-p$ momentum sensitivity of cross sections for such reactions.
This is in contrast with previous $D$-state studies, e.g. \cite{Joh67,Del74}, that
focused on the effects of the D-state component of the reaction vertex $V_{np}|\phi_0
\ra$ in the DWBA amplitude. The conclusions there, for low energy reactions, are that
DWBA cross sections and vector analysing powers are insensitive to the deuteron
$D$-state, the primary sensitivity being on the tensor polarization observables.

In this paper we develop exact adiabatic model $(d,p)$ reaction calculations that use
energy-independent nonlocal nucleon optical potentials and that include the deuteron
$D$-state. We derive formal expressions and calculate the nonlocal deuteron channel
potential ${\cal U}_{dA}$ and the corresponding $d-A$ distorted waves. The effects on
$(d,p)$ cross sections are discussed and compared with those obtained from the earlier,
approximate lowest-order local model. Our key findings, applied to a $^{26}$Al target,
were presented in Ref. \cite{Bai16} and focused on the sensitivity of $\la T_{np}\ra_V$
and the corresponding $(d,p)$ cross sections to high $n-p$ momenta, which is different
between NN models. It was shown that, in some cases, cross sections can change significantly
with different choices of deuteron wave function, and that these changes correlate with
the $D$-state component. Here, we present full details of the model calculations and
extend the model's application to include $^{40}$Ca and $^{28}$Si targets. For $^{28}$Si
we explore a range of neutron separation energies and different orbital angular momentum
transfers. We restrict ourselves to low-energy $(d,p)$ reactions, relevant to ISOL
facilities, where spin-orbit terms of the nucleon optical potentials and finite-range
effects of the transition interaction can be neglected, allowing a clearer evaluation of
the $D$-state effects.

In Sec. II we review the role of the $D$-state on the $d-A$ distorted wave and cross sections
within the standard (local) ADWA and in the DWBA with the Watanabe folding model $d-A$
potential. In Sec. III we then present the formalism for the nonlocal deuteron adiabatic
potential. In Sec. IV we compare the present results, made in a lowest-order approximation,
to those obtained in the local model proposed in Refs. \cite{Tim13a,Tim13b}. In Sec. V we
present the present exact calculations for several targets, focusing on how $D$-state effects
evolve with the separation energy and orbital angular momentum of the transferred neutron.
Implications for extracted spectroscopic factors are discussed for a specific reaction of
astrophysical interest, on a $^{26}$Al target. Conclusions are drawn in Sec. VI. Other
relevant details are presented in an Appendix.

\section{D-state in ADWA with local nucleon optical potentials}
Before presenting our nonlocal potential plus $D$-state $(d,p)$ model we comment on $D$-state
contributions to standard local ADWA calculations. The ADWA includes deuteron break up effects
through a three-body ($A+n+p$) description of the deuteron channel. The $(d,p)$ transition
amplitude in the three-body model is
\begin{equation}
T_{(d,p)} = \sqrt{C^2S} \la {\mathlarger{\chi}}^{(-)}_p \phi_n |V_{np}|\Psi^{(+)}_d \ra,
\end{equation}
where $\Psi^{(+)}_d (\ve{R},\ve{r})$ is the deuteron channel three-body wave function, $\ve{R}$
is the vector separation of the deuteron and the target, $\ve{r}$ the neutron-proton separation
and $V_{np}$ is the  neutron-proton interaction in the deuteron. Following the convention used
in Ref. \cite{Tim13b}, we define $\ve{r}=\ve{r}_n - \ve{r}_p$ and $\ve{R}=-(\ve{r}_n+ \ve{r}_p)
/2$ where $\ve{r}_n$ and $\ve{r}_p$ are the positions of the incident neutron and proton relative
to the mass $A$ target. Thus, with this convention, $\ve{r}_n=\ve{r}/2-\ve{R}$ and $\ve{r}_p=
-\ve{r}/2-\ve{R}$. In the final state, the proton channel distorted wave ${\mathlarger{\chi}
}^{(-)}_p$ is a function of $\ve{R}_p$, the position of the outgoing proton relative to the
product nucleus $B$$(=A+n)$. $\phi_n$ is the normalised bound-state wave function of the
transferred neutron in the final state (more generally, the neutron overlap function) and
$C^2S$ is its spectroscopic factor.

The ADWA makes use of a Weinberg states expansion of $\Psi^{(+)}_d$ \cite{JT}, valid for
$\ve{r}$ values within the range of $V_{np}$, as required to evaluate $T_{(d,p)}$. It was
shown \cite{Pan13} that with this basis, the transition amplitude converges rapidly and
only the first Weinberg state needs to be retained. With this approximation $\Psi^{(+)
}_d(\ve{R},\ve{r}) \rightarrow {\mathlarger{\chi}}^{(+)}_d (\ve{R}) \phi_0 (\ve{r}) $
and $T_{(d,p)}$ in the ADWA is
\begin{equation}
T_{(d,p)} = \sqrt{C^2S} \langle {\mathlarger{\chi}}^{(-)}_p \phi_n |V_{np}|
{\mathlarger{\chi}}^{(+)}_d \phi_0 \rangle,
\end{equation}
where ${\mathlarger{\chi}}^{(+)}_d (\ve{R})$ describes the center-of-mass distortion
of the incident $np$-pair in the presence of deuteron breakup effects. When the
nucleon-target potentials $U_{NA}$ $(N=n,p)$ are local, ${\mathlarger{\chi}}^{(+)
}_d (\ve{R})$ is calculated from the adiabatic distorting potential
\beq
U_{dA} = \la \phi_1 | U_{pA}(-\ve{r}/2-\ve{R}) + U_{nA}(\ve{r}/2-\ve{R}) |\phi_0 \ra.
\eeqn{adpot}
Because of the short range of $\phi_1$, of Eq. (\ref{phi1}), the main contributions
in Eq. (\ref{adpot}) come from values $r \approx 0$. Letting $r\rightarrow 0$ connects
the Weinberg states technique and the earlier Johnson-Soper adiabatic formalism
where, assuming a zero-range $V_{np}$, $U^{\rm JS}_{dA}(R) = U_{nA}(R)+ U_{pA}(R)$
\cite{JS} and, in which limit, the adiabatic potential is seen to be independent of
details of the assumed deuteron wave function. As is now shown, full calculations of
the central terms of $U_{dA}$ of Eq. (\ref{adpot}), with different realistic deuteron
wave functions and local nucleon-target interactions, give very similar results and
show essentially no sensitivity to the $D$-state.

Below we show $U_{dA}$ for the deuteron wave function of the Argonne $V_{18}$ (AV18) NN
interaction \cite{AV18}, with both $S$- and $D$-state components. In the presence of the
$D$-state, we write the deuteron ground state wave function $\phi_0^{M_d}$, with angular
momentum projection $M_d$, as
\beq
\phi_0^{M_d}(\ve{r}) = \sum_{l_d \lambda_d \sigma_d} (l_d \lambda_d s_d \sigma_d|J_d M_d)
\frac{u_{l_d}(r)}{r} Y_{l_d}^{\lambda_d}(\hat{\ve{r}})\chi_{s_d}^{\sigma_d}, \eol
\eeqn{dwf}
with $l_d\lambda_d, s_d\sigma_d$ the orbital and spin angular momenta and their projections
coupled to $J_d\ (=1)$, $\chi$ is the $np$ spinor and the $u_{l_d}(r)$ are the deuteron $S$-
and $D$-state radial wave functions. The vertex function $V_{np} \phi_0^{M_d}$ has an identical
form but with the $u_{l_d}(r)$ replaced by short-ranged radial vertex functions $v_{l_d}(r)$.

\begin{figure}[t]
\includegraphics[scale=0.30]{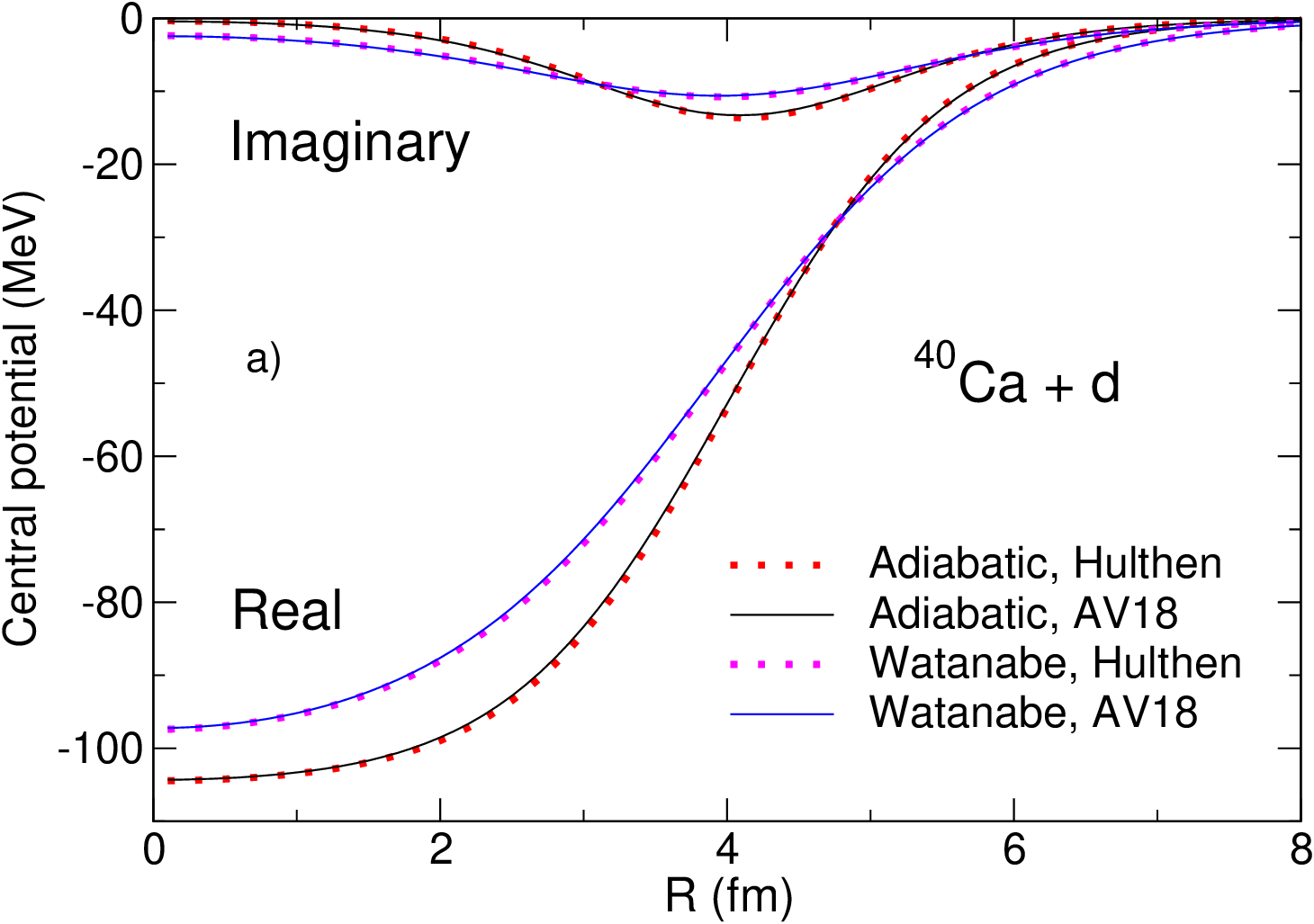}

\vspace{0.8cm}
\includegraphics[scale=0.30]{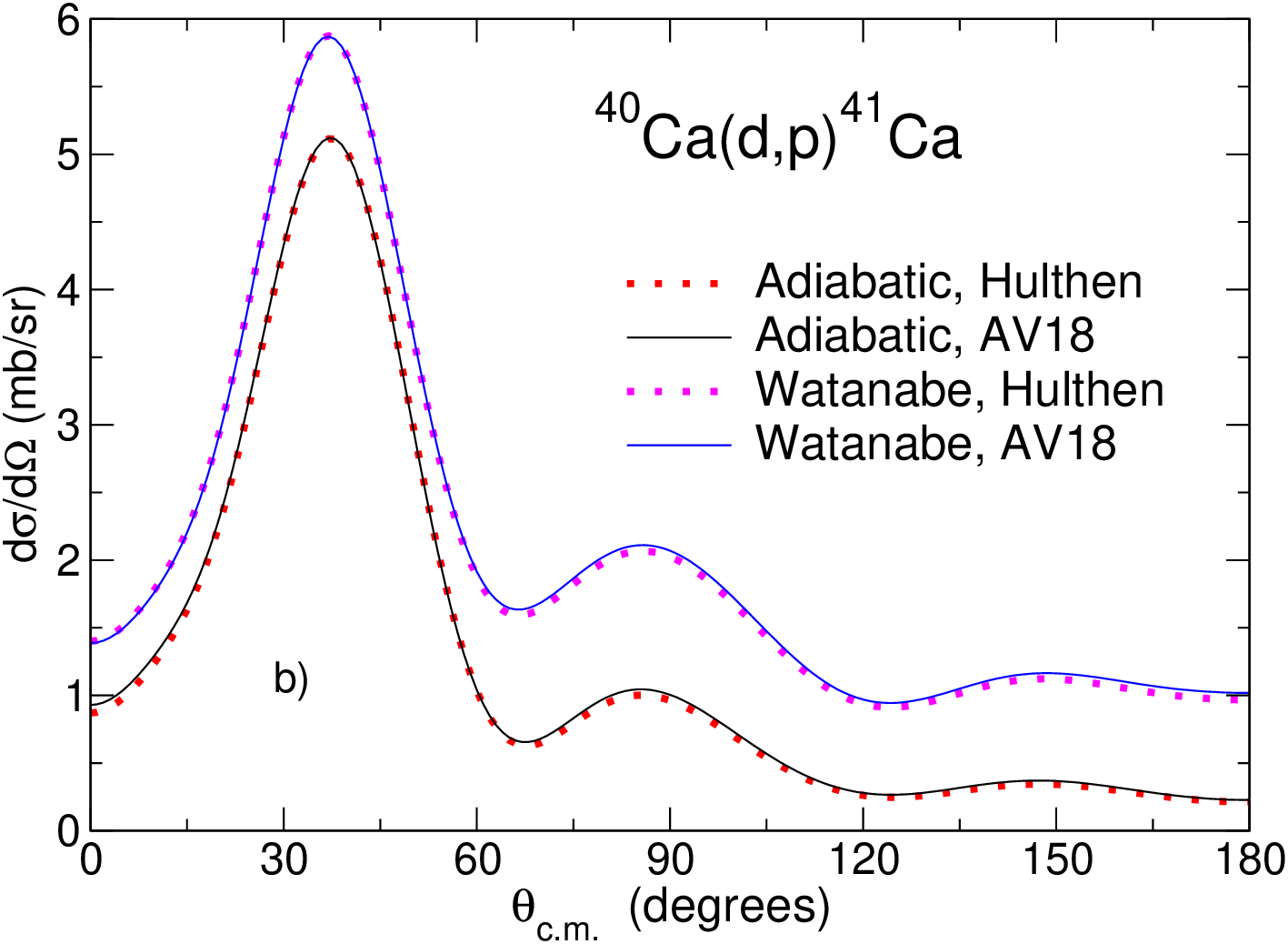}
\caption{($a$) Adiabatic and Watanabe potentials for $d-^{40}$Ca at $E_d = 10$ MeV, and ($b$)
$^{40}$Ca$(d,p)^{41}$Ca differential cross sections for $E_d = 10$ MeV, using the Chapel Hill 89
phenomenological local optical potential for two NN potential models: (i) the central Hulth\'en
potential with an $S$-state deuteron (dots) and (ii) the realistic AV18 potential and deuteron
with both the $S$- and $D$-states (solid lines).}
\label{fig:1}
\end{figure}

The calculated adiabatic potential $U_{dA}$ for the $d-^{40}$Ca system at $E_d = 10$ MeV,
using the Chapel Hill 89 (CH89) phenomenological local optical potential for the $U_{NA}$
\cite{CH89}, is shown in Fig. \ref{fig:1}a. For comparison, the $U_{dA}$ calculated using
the $S$-state Hulth\'en wave function is also shown. The two potentials are very similar,
as are the corresponding $^{40}$Ca$(d,p)^{41}$Ca cross sections, presented in Fig.
\ref{fig:1}b. All of the NN potential models used in Ref. \cite{Bai16} lead to this
same conclusion.
We add that there is also negligible $D$-state sensitivity in the central terms of the
deuteron-target interaction and the transfer reaction cross sections in the no-breakup
limit of the $d-A$ scattering - the Watanabe folding model \cite{Watanabe} - when the
distorting potential is
\beq
U^{\rm Wat}_{dA} = \langle \phi_0 |  U_{pA}(-\ve{r}/2-\ve{R}) + U_{nA}(\ve{r}/2-\ve{R})
|\phi_0 \rangle.\ \ \eeqn{Watanabe}
The calculated Watanabe potentials are also shown in Fig. \ref{fig:1}a for both the Hulth\'en and
AV18 cases. The $D$-state and NN-model insensitivity of the corresponding (DWBA) cross
sections is shown in Fig. \ref{fig:1}b.

In the cross section calculations above, and throughout this paper, we use the zero-range
approximation, $D(\ve{r}) = \langle \ve{r} | V_{np}| \phi_0 \rangle \approx D_0 \delta(
\ve{r})$ to the transition interaction when calculating $T_{(d,p)}$. The volume integrals
$D_0$ are determined for each NN model, and are given in Table \ref{tab:wfscomp} for the
models used here. The zero-range approximation to $T_{(d,p)}$ is very accurate for
reactions of low energy deuteron beams, where finite-range corrections are small, e.g.
\cite{Ngu10}. Hence, we compute the transition amplitudes
\begin{equation}
T_{(d,p)} = D_0 \sqrt{C^2S} \langle {\mathlarger{\chi}}^{(-)}_p \phi_n |
{\mathlarger{\chi}}^{(+)}_d \rangle.
\end{equation}

In the following we calculate $T_{(d,p)}$ when both ${\mathlarger{\chi}}^{(-)}_p$ and
${\mathlarger{\chi}}^{(+)}_d $ are generated by nonlocal potential models - the latter
including realistic deuteron wave functions and $D$-state effects through the adiabatic
model entrance channel effective interaction.

\section{The Nonlocal scattering problem}
In this section we describe the computations of the distorted waves ${\mathlarger{\chi}
}^{(-)}_p$ and ${\mathlarger{\chi}}^{(+)}_d $ for the ADWA calculations in the nonlocal
model case. They satisfy the inhomogeneous Schr\"odinger-like equation, with $\alpha =
p$ or $d$,
\beq
(T_\alpha + U_c(R)&-& E) {\mathlarger{\chi}}_{\alpha}(\ve{R})  =
\eol & -&\int d\ve{R}^{\prime} \ {\cal U}_{\alpha}(\ve{R}, \ve{R}^{\prime})
\ {\mathlarger{\chi}}_{\alpha}(\ve{R}^{\prime}),
\eeqn{Schdist}
where $E$, $T_\alpha$ and $U_c$ are the center of mass energy, kinetic energy
operator and Coulomb interaction and ${\cal U}_{\alpha}(\ve{R}, \ve{R}^{\prime})$
is the nonlocal nuclear potential in channel $\alpha$. Throughout this work we
adopt an energy-independent Perey-Buck parameterization of the nonlocal nucleon-target
potentials. Equation (\ref{Schdist}) is solved iteratively, following partial wave
decomposition, from a trial complex and local starting potential $U_\alpha(R)$,
namely
\beq
&&(T_\alpha + U_\alpha(R)+U_c(R) - E) {\mathlarger{\chi}}^{(i+1)}_{
\alpha}(\ve{R})  = \eol &-&\int \!d\ve{R}^{\prime} \left[{\cal U}_{\alpha}(
\ve{R},\ve{R}')-U_\alpha({R})\delta(\ve{R}-\ve{R}')\right] {
\mathlarger{\chi}}^{(i)}_{\alpha}(\ve{R}^{\prime}) \nonumber
\eeqn{Sch2}
with $\mathlarger{\chi}^{(0)}_{\alpha}$ the solution of the homogeneous equation
with the appropriate scattering boundary conditions. Our treatment of the proton
and deuteron channel nonlocal potentials is described in detail below.

\subsection{Proton Channel}
The nonlocal interaction in the proton channel describes the motion of the
outgoing proton with respect the resultant nucleus $B$$(=A+n)$. This is of
Perey-Buck type, i.e.
\begin{equation}
{\cal U}_{pA}(\ve{R}_p,\ve{R}'_p) = H(|\ve{R}_p-\ve{R}'_p|) U_{pA}\bigg(
\frac{|\ve{R}_p+\ve{R}'_p|}{2}\bigg), \label{eq:PB}
\end{equation}
with $H$ a normalized Gaussian (in 3 dimensions) with a range $\beta$,
\beq
H(x)=(\sqrt{\pi}\beta)^{-3}\exp(-x^2/\beta^2). \eeq
The potential form factors $U_{NA}$ are complex with conventional Woods-Saxon
real parts and surface-peaked derivative Woods-Saxon imaginary parts. This
nonlocal interaction is used directly in the source term of the inhomogeneous
equation, Eq. (\ref{Schdist}). Thus, the proton channel partial wave functions
${\mathlarger{\chi}}^{J}_{L}$, with $\ve{J}=\ve{L}+\ve{s}_p$, $s_p=1/2$,
satisfy
\beq
&\left(T^{(L)}_{p} + U_c(R_p) - E_p\right) {\mathlarger{\chi}}^{J}_{L}(R_p)
= \qquad\qquad\qquad \eol & -R_p \int_0^\infty dR_p^{\prime}\,\, R'_p\, {\cal U}^{
(p)}_{L} (R_p,R_p^{\prime}) {\mathlarger{\chi}}^{J}_{L}(R_p')
\eeqn{protondist}
with
\beq
T_\alpha^{(L)} = -\frac{\hbar^2}{2m_\alpha}\left[\frac{d^2}{dR^2_\alpha}-
\frac{L(L+1)}{R^2_\alpha}\right] \label{kinen}
\eeq
and $m_\alpha$ is the reduced mass in channel $\alpha$. The potential kernel is
\beq
{\cal U}^{(p)}_{L} (R_p,R_p') &=& 2 \pi \int_{-1}^1 d\mu \,\,  P_L(\mu) H(|\ve{R}_p
-\ve{R}_p'|) \eol &\times& U_{pA}\bigg(\frac{|\ve{R}_p+\ve{R}_p'|}{2}\bigg),
\eeq
where $\mu= \ve{R}_p \cdot \ve{R}'_p/R_p R'_p$ and $P_L$ is the Legendre polynomial
of order $L$. With the neglect of spin-orbit interactions, as assumed here, the
$J=L\pm1/2$ channel distorted waves are of course identical.

In all of the nonlocal ADWA calculations presented, the exact solutions of
Eq. (\ref{protondist}) are read into the transfer reactions code {\sc twofnr}
\cite{twofnr}. Comparisons between such exact solutions and those from a
phase-equivalent model can be found in Ref. \cite{Tit14}.

\subsection{Deuteron Channel}
The nonlocal deuteron channel potential ${\cal U}_{dA}(\ve{R},\ve{R}')$ is
constructed using nonlocal nucleon-target optical potentials with the Perey-Buck
form of Eq. (\ref{eq:PB}). The formal expression for the Johnson-Tandy adiabatic
model potential ${\cal U}_{dA}(\ve{R},\ve{R}')$ in terms of the nucleon potentials
${\cal U}_{NA}(\ve{R}_N,\ve{R}'_N)$ is given by Eq. (12) of Ref. \cite{Tim13b} - a
folding-type integral where the arguments of the nucleon optical potentials are
reexpressed in terms of the deuteron channel variables $\ve{R}$, $\ve{R}'$ and
$\ve{r}$. Here, we take the target mass $A$ to be infinitely large in the more
general expression of Ref.\cite{Tim13b}. The partial wave form of ${\cal U}_{dA}
(\ve{R},\ve{R}')$ is more complicated when the $D$-state is present and the
required expansion is most easily achieved by use of the variables $\ve{R}$
and $\ve{S}=\ve{R}-\ve{R}'$. We obtain
\beq
&&{\cal U}_{dA}^{M_d M'_d}  (\ve{R},\ve{S}) = 8 H(2\ve{S}) \int\! d\ve{x}\,
\phi_1^{*{M_d}}(\ve{x}-2\ve{S})\eol &&  \times \left[ U_{nA}(\frac{\ve{x}}{2}
-\ve{R}) + U_{pA}(\frac{\ve{x}}{2}-\ve{R}) \right]\phi_0^{M'_d}(\ve{x}) ,
\eeqn{eq:Uda}
where $M_d, M'_d$ are the projections of the intrinsic angular momentum $J_d$ (=1)
of the deuteron referred to the incident beam direction.

We multipole expand the vertex function $\phi_1$ and the nonlocal nucleon-target
potential formfactors in terms of the vectors $\ve{x}$, $\ve{S}$ and $\ve{R}$,
and separate their radial and angular components. After summation over angular
momentum projections, ${\cal U}_{dA}$ is the operator in deuteron spin space
\beq
{\cal U}_{dA}(\ve{R},\ve{S})& = & 4\pi
\sum_{l_1 l_2 a} v_{l_1 l_2}^a (R, S) \sum_{\alpha} (-1)^{a-\alpha} {\cal T
}_{a-\alpha} \eol &\times& \bigg[{Y}_{l_1}(\hat{\ve{R}}) \otimes {Y}_{l_2}
(\hat{\ve{S}}) \bigg]_{a \alpha},
\eeqn{Udafinal}
where ${\cal T}_{kq}$ is the irreducible tensor operator of rank $k$ in the
space of spin $J_d$ with matrix elements
\beq
\la J_d M_d|{\cal T}_{kq}|J_d M_d'\ra = \hat{k} (J_d M_d' kq|J_d M_d) .
\nonumber
\eeq
The functions $v_{l_1 l_2}^a$ contain all information on the deuteron
wave function and nonlocal potential form factors, details of which
are presented in an Appendix. Further expansion of $v^a_{ l_1 l_2}$
and ${Y}_{l_2}(\hat{\ve{S}})$, now with respect to $\ve{R}$ and $\ve{R}'$,
then derives the required (radial variables) kernel of ${\cal U}_{dA}$,
\begin{widetext}
\begin{eqnarray}
\label{eq:sourceterm}
{\cal U}^{J}_{L' L''} (R,R') &=& \ 2\pi\displaystyle\sum\limits_{\substack{a
l_1 l_2}} \sum_{\substack{{\cal L}{\cal L}'\\\tau+\eta=l_2}} \hat{a}^2
\hat{J}_d \hat{l}_1 \hat{l}_2 \hat{{\cal L}}^2 \hat{{\cal L}'}  \left[\frac{
(2l_2 + 1)!}{(2\tau)!(2\eta)!} \right]^{\frac{1}{2}}(\tau0{\cal L}0|{\cal L}'0)
(\eta0{\cal L}0|L''0)(l_10{\cal L}'0|L'0)  W({\cal L}' {\cal L} l_2 \eta;\tau
L'') \eol &\times& W(L'' {\cal L}' a l_1 ;l_2 L')  W(L' a J J_d;L'' J_d) R^{\tau}
R'^{\eta}   \int^1_{-1} \frac{v^a_{ l_1 l_2} \left(R,\left|\ve{R}-\ve{R}'\right|
\right)}{\left|\ve{R}-\ve{R}'\right|^{l_2}} P_{{\cal L}} (\mu) \ d\mu.
\end{eqnarray}
\end{widetext}
Here, in standard notations, $\hat{x}=\sqrt{2x+1}$, $W$ is the Racah coefficient and
$\mu=\ve{R}\cdot\ve{R}'/RR'$.
These nonlocal kernels, ${\cal U}^{J}_{L' L''} (R,R')$, enter the Schr\"odinger-like
equation for the deuteron distorted waves ${\mathlarger{\chi}}^{J}_{L'L}(R)$
for total angular momentum $J$. Explicitly,
\begin{eqnarray}
\label{eq:Schdistfinal}
(T^{(L')}_d &+& U_c(R) - E_d) {\mathlarger{\chi}}^{J}_{L'L}(R)\eol  =-&R&
\displaystyle\sum\limits_{L''} \int^{\infty}_0 dR^{\prime} R^{\prime} \ {\cal U
}^{J}_{L' L''} (R,R') {\mathlarger{\chi}}^{J}_{L''L}({R}^{\prime})
\end{eqnarray}
where $T^{(L')}_d$ is given by Eq. (\ref{kinen}).

The central terms of ${\cal U}_{dA}$, corresponding to $a=0$, are of course diagonal in
the orbital angular momentum quantum number and
\begin{eqnarray}\label{eq:a0st}
&&{\cal U}^{J}_{L'L''}(R,R') = 2\pi\delta_{L'L''}  \displaystyle\sum\limits_
{\substack{l{\cal L}{\cal L}'\\\tau+\eta=l}}  \frac{\hat{{\cal L}}^2 \hat{{\cal L}'}
\hat{l}}{\hat{L'}^2} \left[\frac{(2l + 1)!}{(2\tau)!(2\eta)!} \right]^{\frac{1}{2}}\eol
&&\times(\tau0{\cal L}0|{\cal L}'0) (\eta0{\cal L}0|L'0)  (l0{\cal L}'0|L'0) W({\cal L}'
{\cal L} l \eta;\tau L')\eol &&\times R^{\tau} R'^{\eta} \int^1_{-1} \frac{v^0_{ l l}
\left(R,\left|\ve{R}-\ve{R}'\right|\right)}{\left|\ve{R}-\ve{R}'\right|^{l}} P_{{\cal L}}
(\mu) \ d\mu.
\end{eqnarray}
The smaller potential terms with $a=2$, of tensor character, arise only when the
deuteron $D$-state is included.

\section{Lowest-order approximation to the nonlocal model}
Before calculating these ${\cal U}_{dA}$ exactly, we investigate the potential calculated
in an approximation, called the lowest-order (LO) limit. Here, due to the short range of
$\phi_1$ in the folding integral, Eq. (\ref{eq:Uda}), we replace
\begin{equation}\label{eq:lolim}
U_{NA}(|\ve{x}/2-\ve{R}|) \rightarrow U_{NA}({R})
\end{equation}
as was also used in \cite{Tim13a,Tim13b}, there assuming an $S$-wave deuteron. This limit
calculates the leading-order contributions to the potential and the $(d,p)$ cross section
and provides insight into the nature of the derived entrance channel interaction. It was
shown previously \cite{Tim13b}, in this limit and within the local-energy approximation
(LEA), that the adiabatic potential is local and that this local potential, $U_{dA}^{\rm
loc}$, solves the transcendental equation
\beq
U^{\rm loc}_{dA} = M_0 U_{d}(R) \exp\left[-\frac{ m_d \beta_d^2} {2\hbar^2} (E_d-U^{
\rm loc}_{dA}- U_c)\right],\ \
\eeqn{ulocd}
where $m_d$ is the deuteron channel reduced mass, $\beta_d$ is the effective deuteron
nonlocality range \cite{Tim13b} and $M_0$ is the zeroth-order moment of the nonlocality
factor
\beq
M_0 = \int\! d\ve{s} \int \!d\ve{x} H(s) \phi_1(\ve{x}-\ve{s})\phi_0(\ve{x}).
\eeqn{M0}
In Eq. (\ref{ulocd}) and below, $U_{d}(R)= U_{nA}(R)+U_{pA}(R)$.

Further, it was shown, for $Z=N$ nuclei, that $U_{dA}^{\rm loc}$ can be constructed
from phenomenological local nucleon optical potentials that describe elastic scattering
at an energy shifted by $\Delta E$ from the usually assumed value, $E_d/2$, of half
the incident deuteron energy. Through $M_0$, this shift $\Delta E$ is determined by
$\la T_{np}\ra_V$ of Eq. (\ref{Tnp}). Here, we find that Eq. (\ref{ulocd}) and its
shifted-energy solution remain valid when a deuteron $D$-state is present. However,
the values of $\la T_{np}\ra_V$, $M_0$ and $\Delta E$ are strongly affected by the
presence of the $D$-state component. For example, Table \ref{tab:wfscomp} shows that
the $\Delta E$ values for the neutron, from NN models with a $D$-state component,
typically span an interval from 38--75 MeV, significantly larger than the $\Delta E$
of 32 MeV from the $S$-wave Hulth\'en NN model.
\begin{table}[h]
\centering
\begin{tabular}{l cccccc }
\hline\hline
NN Model \qquad&\ \ $P_D$\ \  & \ \ $D_0$ \ \  & $\ \ \langle\ T_{np}\rangle_V
\ \ $ & $\Delta E$\\ & $\%$&MeV fm$^{\frac{3}{2}}$&MeV&MeV\\
\hline
Hulth\'{e}n       &  0   &  $-$126.15  & 106.6 & 31.7 \\
Reid soft core    & 6.46 &  $-$125.19  & 245.8 & 75.3 \\
Argonne V18       & 5.76 &  $-$126.11  & 218.0 & 66.8 \\
CD-Bonn           & 4.85 &  $-$126.22  & 112.5 & 37.6 \\
\hline\hline
\end{tabular}
\caption{D-state percentages $P_D$, volume integrals $D_0$ of the transfer
vertex $D(\ve{r})$, and short-ranged $n$-$p$ kinetic energy $\langle\ T_{np}
\rangle_V $ for different NN-model interactions used here. The neutron energy
shifts $\Delta E$ are calculated for the d+$^{40} $Ca system at $E_d=11.8$ MeV
and are computed using the lowest-order methodology discussed here. The proton
energy shifts are larger by the Coulomb energy, $\approx$6.8 MeV for $^{40}$Ca
+ p scattering. Details can be found in sections IV.A and IV.B of Ref.
\cite{Tim13b}.} \label{tab:wfscomp}
\end{table}

Below we will present, as typical, calculations for the phenomenological
AV18 NN potential. The other NN potentials studied, see Table \ref{tab:wfscomp}
and also Table I of Ref. \cite{Bai16}, give both larger and
smaller $\Delta E$. We note also that, if we include only the $S$-wave part
of the AV18 wave function, then the values of both $M_0$ and $\Delta E$ are
very similar to those of the Hulth\'en wave function. Thus, the primary
difference between the AV18 and Hulth\'en model results arises from the
$D$-state component of the wave function.

\begin{figure}[t]
\includegraphics[scale=0.30]{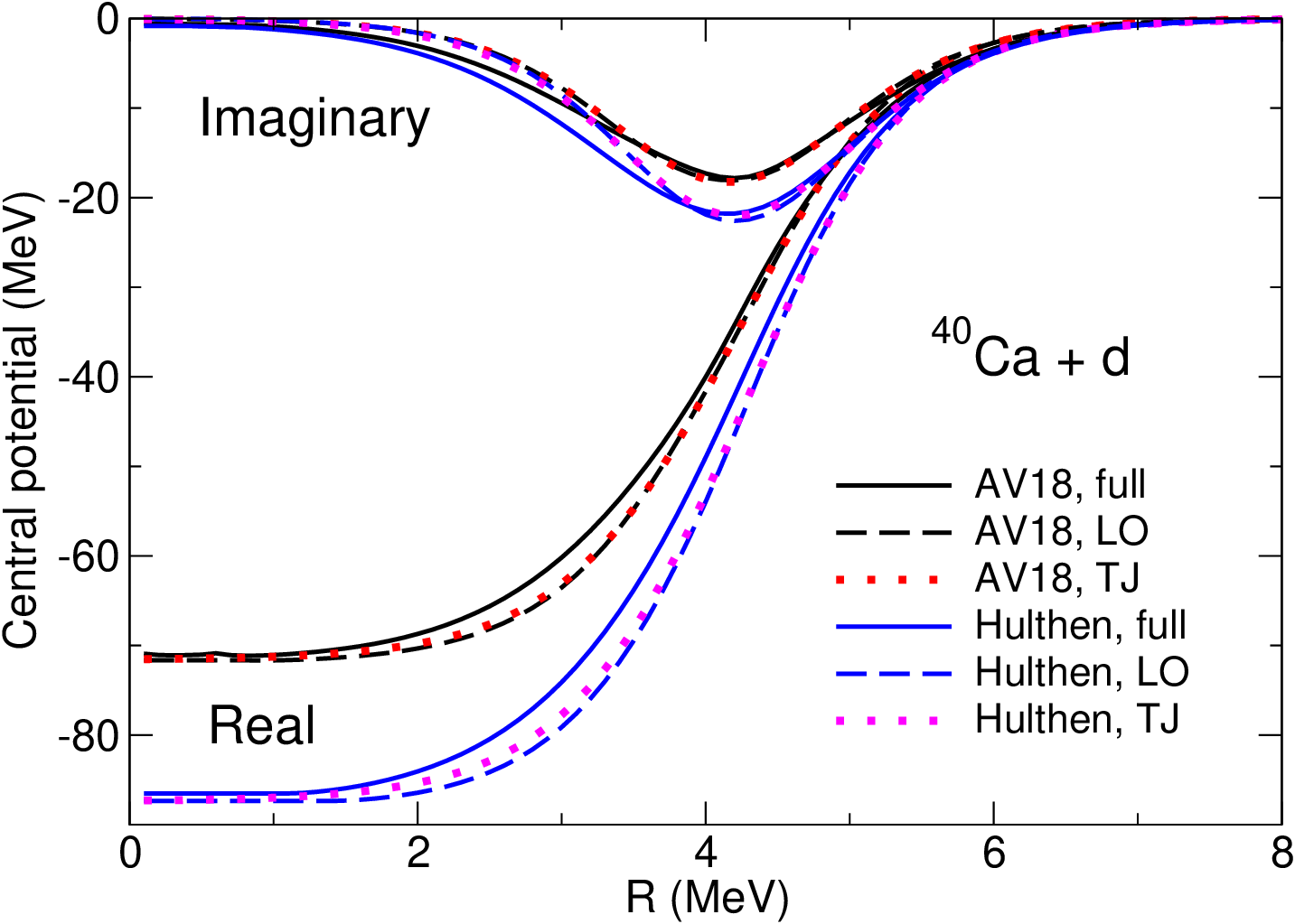}

\vspace{1.0cm}
\includegraphics[scale=0.30]{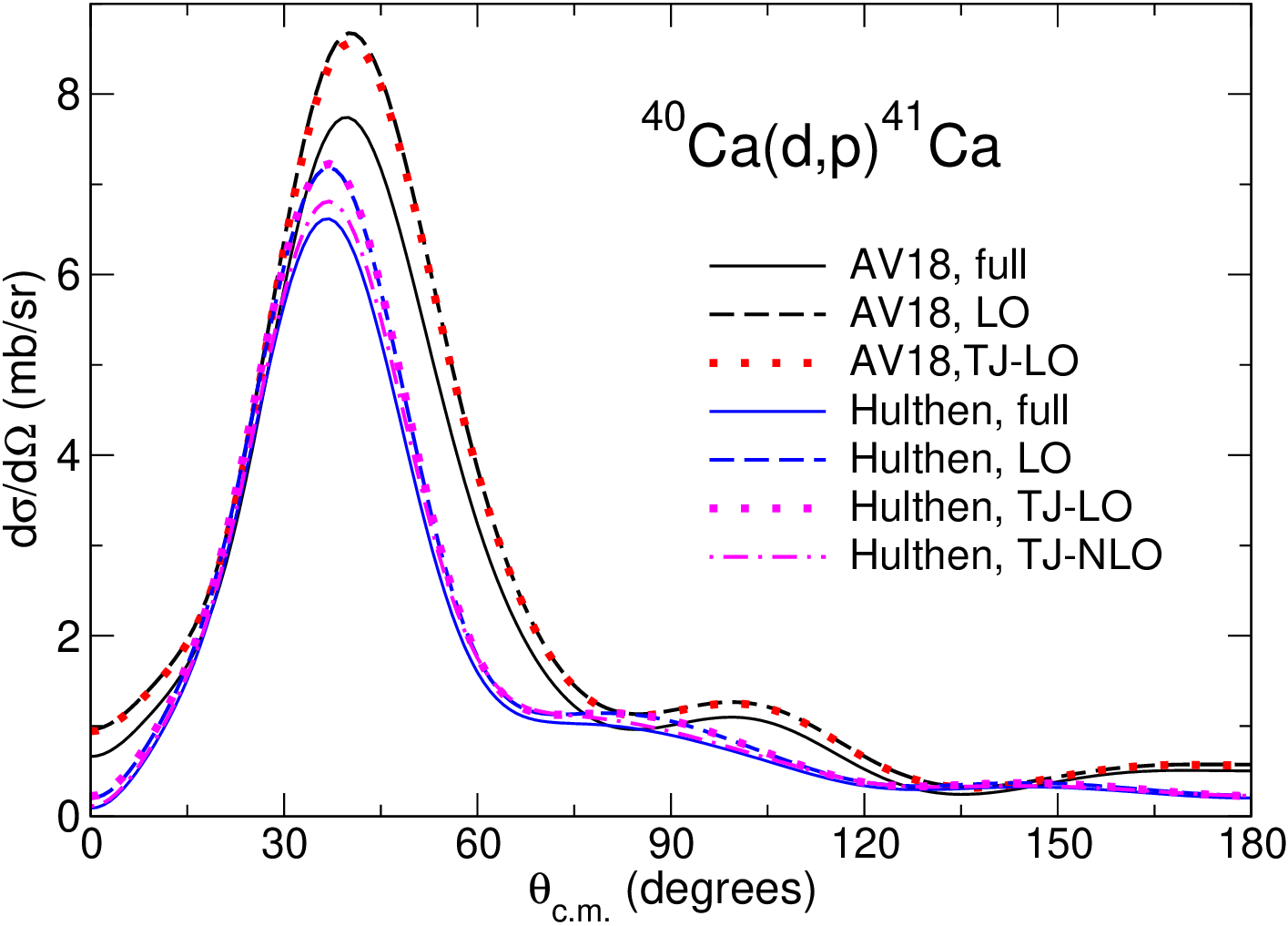}
\caption{($a$) Trivially-equivalent local adiabatic potentials for $d-^{40}$Ca,
and ($b$) the $^{40}$Ca$(d,p)^{41}$Ca differential cross sections obtained from
full (solid lines) and lowest-order (dashed lines) calculations at $E_d = 11.8$
MeV. Calculations assume the $S$-wave Hulth\'en potential (blue) and the more
realistic AV18 NN potential with both $S$- and $D$-states (black). These are
compared to the lowest-order TJ results for the same NN potentials. For the
Hulth\'en potential, the TJ cross sections in next-to-leading order are also
shown by the dashed-dotted (magenta) curve.}
\label{fig:2}
\end{figure}

In the LO limit, the nonlocal potential of Eq. (\ref{Udafinal}) simplifies to
\begin{equation}
\label{eq:3}
{\cal U}^{({\rm LO})}_{dA} (R, \ve{S}) = \sqrt{4\pi}{U}_d(R) \sum_{a \alpha}
\nu_a (S) (-1)^{a-\alpha} {\cal T}_{a - \alpha} {Y}^{\alpha}_{a}(\hat{\ve{S}}),
\nonumber
\end{equation}
where
\beq
\nu_a(S) &=& 8 H(2S) \frac{\hat{J}_d}{\hat{a}} \sum_{l_d l'_d} \hat{l}_d'
W(a l_d J_d s_d;l'_d J_d) \eol &\times&  \int_0^\infty dx \, x\, \tilde{
\upsilon}_{l'_d a}^{(l_d)} (x,2S) u_{l'_d}(x) .
\eeq
The central terms ($a=0$) of this LO nonlocal potential are diagonal in $L$
with ${\cal U}^{J}_{L L'}(R,R') = {\cal U}^{J}_{L} (R,R')\delta_{LL'}$,
and
\begin{equation}
{\cal U}^{J}_{L}(R,R')= 2\pi {U_d}(R) \int^1_{-1} \nu_{0} \left(\left|\ve{R}-\ve{R}'
\right|\right) P_{L} (\mu) \ d\mu.\nonumber \label{eq:Va0LO}
\end{equation}
We have solved the Schr\"odinger equation for the distorted waves $\chi_{L}(R)$ in
this limit and then constructed the trivially-equivalent local potentials (TELPs)
\begin{equation}\label{eq:loceq}
U_{\rm TELP} (R) = \frac{R\int dR^{\prime}{R^{\prime}} \ {\cal U}_{L} (R,R')
{\mathlarger{\chi}}_{L}({R}^{\prime})}{{\mathlarger{\chi}
}_{L}(R)}
\end{equation}
which can be compared with the lowest-order adiabatic potentials of the approximation
used by Timofeyuk and Johnson (TJ) in Ref. \cite{Tim13b}. This comparison, for the
$d-^{40}$Ca system at $E_d = 11.8$ MeV, is shown in Fig. \ref{fig:2}. In these and
all subsequent calculations we use the nonlocal nucleon-nucleus potential parameterization
of Giannini and Ricco \cite{GR76}, that we denote GR76. We find that the calculated
TELP are essentially independent of $L$ and differ from $U_{dA}^{\rm loc}$ by no more
than 1$\%$ and 2$\%$ for the AV18 and Hulth\'en potentials, respectively. This confirms
that, in leading order, the adiabatic potentials can also be obtained from local nucleon
potentials by applying an appropriate energy shift. Since this shift is larger when the
deuteron $D$-state is included, being 67 MeV for AV18, compared to 32 MeV for the Hulth\'en
case, the adiabatic optical potentials should be shallower - as confirmed by the direct
calculations shown in Fig. \ref{fig:2}a. The cross sections for the $^{40}$Ca$(d,p)^{
41}$Ca reaction using ${\cal U}_{dA}^{(\rm LO)}$, shown in Fig. \ref{fig:2}b, are
also very close to those obtained with $U_{dA}^{\rm loc}$ of the TJ approach, given
by Eq. (\ref{ulocd}). Including the $D$-state is seen to increase the computed $(d,p)$
cross sections.
\begin{figure}[t]
\includegraphics[scale=0.4]{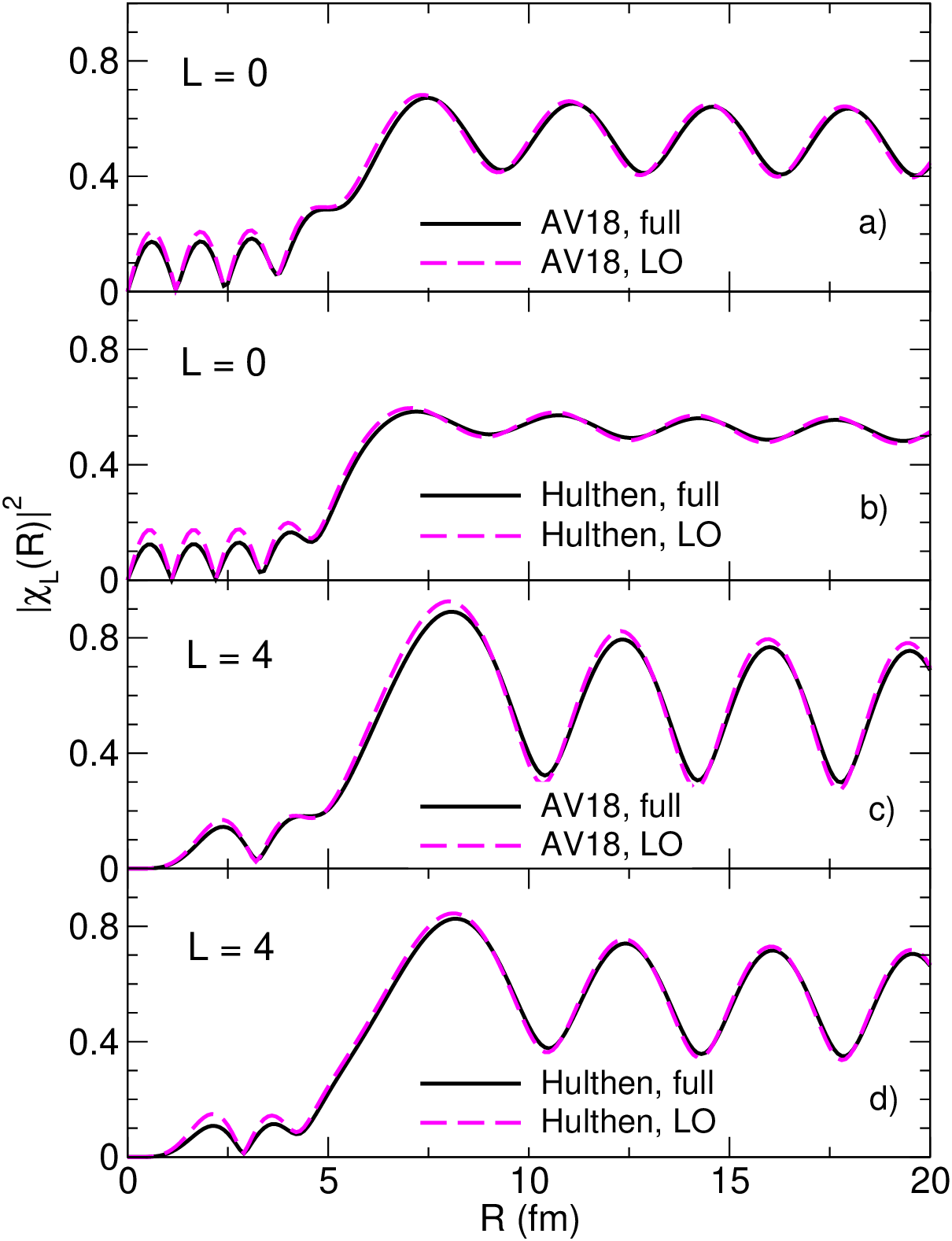}
\caption{The $E_d = 11.8$ MeV, $d-^{40}$Ca distorted waves in $L=0$ ($a,b$) and $L=4$
($c,d$) partial waves, calculated using the AV18 ($a,c$) and Hulth\'en ($b,d$) NN model
potentials. The exact nonlocal calculations are shown by the solid lines while the
approximate, lowest-order (LO) calculations are shown as dashed lines.}
\label{fig:3}
\end{figure}

We note that, by making only the LO approximation of Eq. (\ref{eq:lolim}) to
the full expression for ${\cal U}_{dA}$ (of Eq. (\ref{eq:Uda})) we take into
account all corrections beyond the LEA in the TJ derivation of $U_{dA}^{\rm
loc}$. However, as was shown in \cite{Tim13b}, the first-order correction to
the LEA involves the fourth power of the (small) nonlocality range parameter
$\beta$, suggesting this and higher order corrections will be negligible.
This expectation is indeed confirmed by our comparisons of $U_{dA}^{\rm loc}$
with the exact LO results in Fig. \ref{fig:2}.

\section{Results from full nonlocal calculations}
We now present the results of calculations that compute the exact solutions of
the nonlocal integro-differential problem, as given by Eqs. (\ref{eq:sourceterm})
and (\ref{eq:Schdistfinal}). We note that such solutions could also be approached
by systematic development of the higher-order corrections to the LO model of
the previous section. For example, for the pure $S$-state deuteron case, the
next-to-leading order (NLO) corrections were discussed above and in Ref.
\cite{Tim13a} and, for the Hulth\'en wave function and the $^{40}$Ca$(d,p)^{41}$Ca
reaction we have shown that these differ from the exact calculations by
$\approx$3$\%$, see Fig. \ref{fig:2}b. However, no such systematic development
has yet been carried out when the $D$-state is present.

\subsection{Adiabatic distorted waves and tensor force effects}
The magnitude of the beyond-LO effects in the exact calculations when the
deuteron $D$-state is present are assessed in Fig. \ref{fig:3}. The deuteron
channel partial waves for $L=0$ and $L=4$ are shown for both the Hulth\'en
and AV18 NN potentials. We note that the wave functions of the full calculations
are smaller than the LO results in the nuclear interior, as could be described
by a Perey effect. The oscillations of the wave functions obtained with AV18
are also more pronounced, indicative of a reduced deuteron channel absorption.
The TELPs deduced from these wave functions are presented in Fig. \ref{fig:2}a
showing a complicated dependence of their depths and surface diffuseness on
both the NN interaction model and the presence of higher-order terms in the
presence of the $D$-state.

\begin{figure}[t]
\includegraphics[scale=0.35]{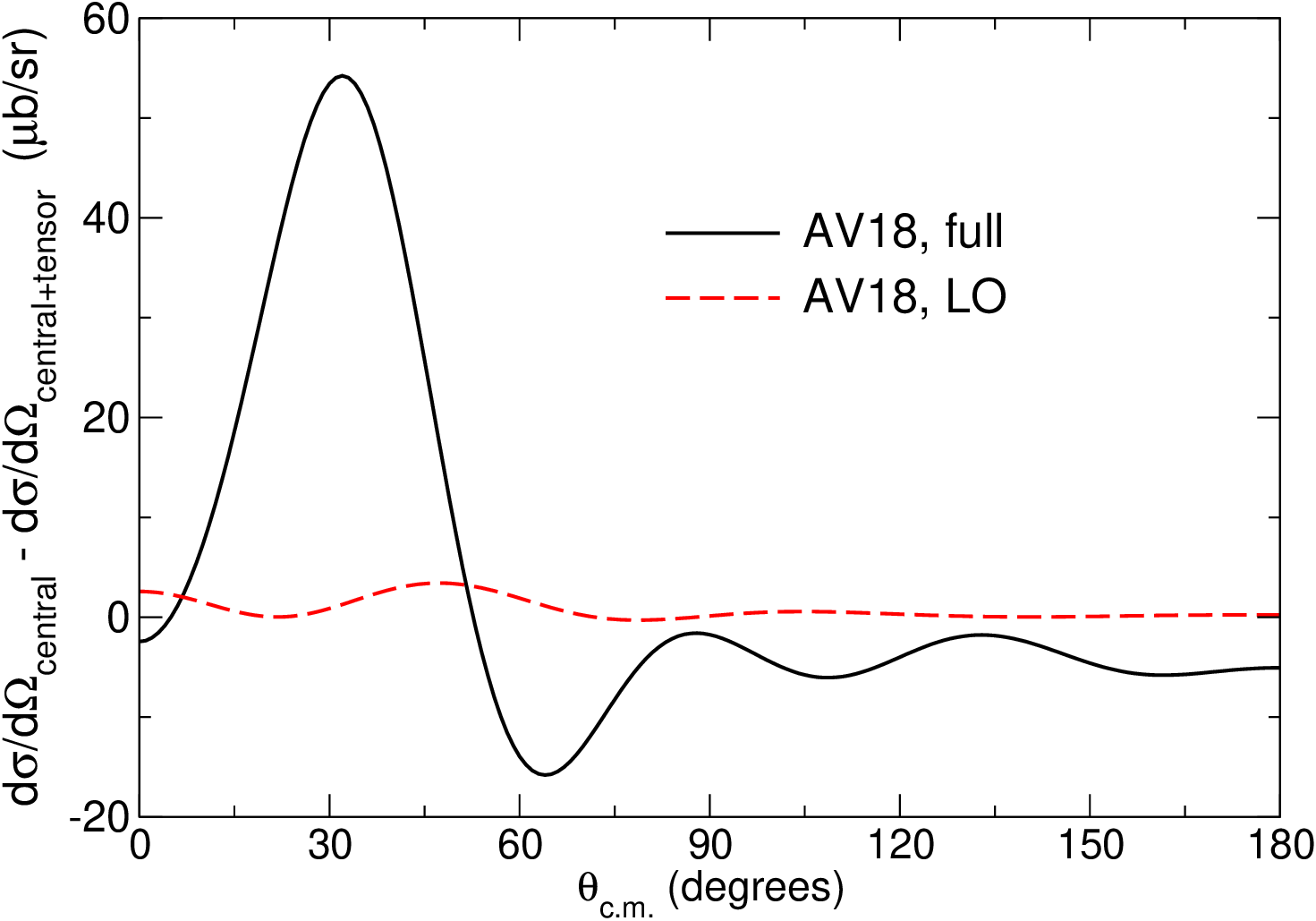}
\caption{Difference between the $^{40}$Ca$(d,p)^{41}$Ca differential cross
sections, for $E_d$ = 11.8 MeV, calculated with only the central terms ($a=0)$
and with the full ($a=0,2$) nonlocal adiabatic potential that includes the
$D$-state generated tensor terms (solid line). The corresponding leading
order (LO) potential results are shown by the dashed line.}
\label{fig:5}
\end{figure}

These wave functions and TELPs from the exact calculations include the small
contributions from the $a=2$ (rank-2 spin-tensor) terms of ${\cal U}_{dA}$,
that generate off-diagonal contributions to the nonlocal potential kernels
${\cal U}^{J}_{L'L}(R,R')$ and the associated deuteron channel partial wave
S-matrices $S^J_{L'L}$. The latter have maximum values of order $10^{-2}$
in the cases studied here. These tensor force effects are included fully in
the exact distorted wave functions input to the transfer reaction calculations.
The importance of these terms, for low energy $(d,p)$ reaction cross sections
was assessed by comparing calculations that include and neglect the $a=2$
contributions to the adiabatic potential. This difference, for the $^{40}
$Ca$(d,p)^{41}$Ca cross sections, in Fig. \ref{fig:5}, is less then 50
$\mu$b/sr and represents a change in the calculated differential cross
sections of 0.6$\%$ or less. Thus, the tensor force effects in the nonlocal
${\cal U}_{dA}$, arising from the deuteron $D$-state, are insignificant
for cross-section-based nuclear spectroscopy and astrophysical studies.

\subsection{$D$-state effect on cross sections: transferred orbital
angular momentum and separation energy dependence}
All calculations above were for the $^{40}$Ca$(d,p)^{41}$Ca(g.s) reaction,
carried out in zero-range approximation. The single-particle bound state
wave function of the transferred $1f_{7/2}$ neutron was obtained in the
standard potential model description, the Woods-Saxon binding potential
having a radius parameter $r_0$ = 1.25 fm, diffuseness $a_0=0.65$ fm and
spin-orbit depth $V_{SO}=6$ MeV. Analogous calculations for the $^{26}$Al$
(d,p)^{27}$Al reaction, populating several final states, were presented in
Ref. \cite{Bai16}. As noted there, the magnitude of the cross section
changes when using realistic deuteron wave functions, with a $D$-state,
depends on the details for the final state of the transferred neutron. We
now discuss a more systematic study of this observed sensitivity.

\begin{figure}
\includegraphics[scale=0.35]{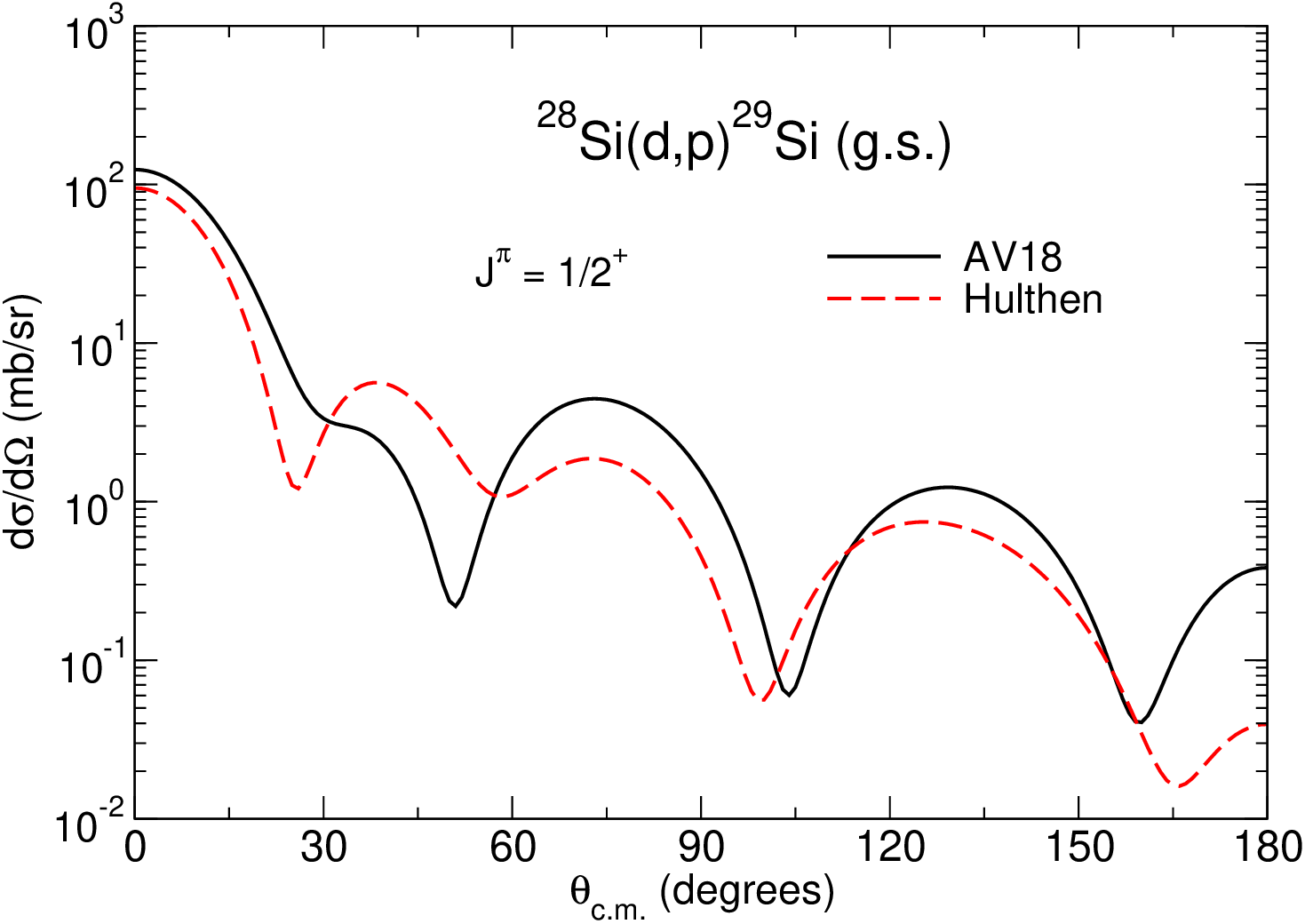}
\caption{\label{fig:snpeak1} The $^{28}$Si$(d,p)^{29}$Si(g.s.) reaction
differential cross sections, at $E_d = 10 $ MeV, obtained using the
${\cal U}_{dA}$ calculated with the $S$-state Hulth\'en and $S+D$-state
AV18 deuteron wave functions.}
\label{fig:6}
\end{figure}

As an example we use the $^{28}$Si$(d,p)^{29}$Si reaction at $E_d=10$
MeV, $^{29}$Si having a rich excitation spectrum with low-lying states
from several single-particle orbitals. We perform calculations using
the GR76 nonlocal nucleon optical potentials and the binding potential
geometry stated above. Fig. \ref{fig:6} shows the predicted $^{29}$Si
ground-state ($J^\pi=1/2^+$) differential cross section calculated in
the $S$-state Hulth\'en and AV18 deuteron cases. The $D$-state is seen
to lead not only to an enhanced cross section in the forward peak but
also to a modified angular distribution at larger angles. So, changes
are complex and not simply a scaling and comparisons with data may
depend sensitively on the available range of measured angles.

\begin{figure}[b]
\bigskip
\includegraphics[scale=0.35]{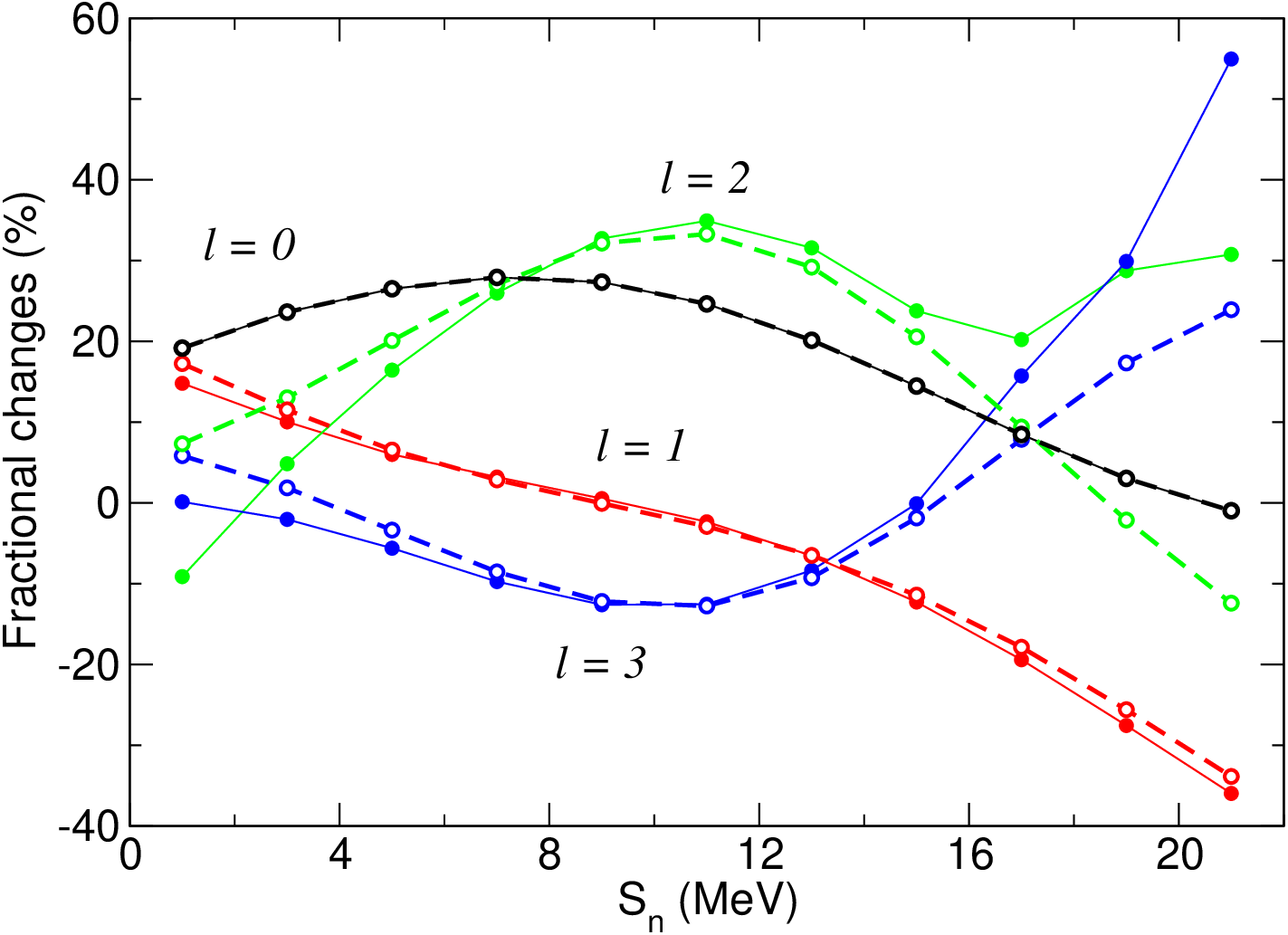}
\caption{\label{fig:7} Fractional changes (as $\%$) in the differential
cross sections (at their first peak) of calculations with the Hulth\'en
and AV18 deuteron wave functions (closed circles and solid lines). Results
are for the $^{28}$Si$(d,p)^{29}$Si reaction. The changes are shown as
functions of the assumed separation energy of the transferred neutron
with orbital angular momenta $l=0,1,2,3$. The changes in the same cross
sections, but at a fixed center-of-mass angle, are also shown (open circles
connected by dashed lines). The fixed angles used were 0, 16, 31 and 44
degrees for $l=0,1,2,3$, respectively.}
\end{figure}

To explore this cross section $D$-state sensitivity further, plus its
dependence on the neutron separation energy, we have performed a series
of calculations with both the Hulth\'en and AV18 wave functions. As above,
these are for the $^{28}$Si$(d,p)^{29}$Si reaction at 10 MeV. Here we have
varied the assumed neutron separation energy between 1 to 21 MeV for four
assumed transitions of different orbital angular momentum, namely: $2s_{
\frac{1}{2}}$ ($l=0$), $2p_{\frac{3}{2}}$ ($l=1$), $1d_{\frac{3}{2}}$
($l=2$) and $1f_{\frac{7}{2}}$ ($l=3$). The fractional changes (as $\%$)
in the differential cross sections (at their first peak) of calculations
with the Hulth\'en and AV18 deuteron wave functions are shown in Fig.
\ref{fig:7}. These ratios depend on the neutron separation energy showing
that inclusion of the $D$-state can result in cross section changes of up
to 30$\%$. The cross section changes for another (fixed) center-of-
mass angle are also presented (see caption to Fig. \ref{fig:7}) showing
that the cross section shapes may also change and that the
peak value ratios may not represent an simple overall scaling. In all
cases the dependence on the neutron separation energy is significant
and changes can reach values of around 50$\%$. Such differences would
certainly affect the interpretation of the experimental data in terms
of a deduced spectroscopic strength.

\begin{table*}[t]
\caption{Deduced spectroscopic factors for the $^{26}$Al$(d,p)^{27}$Al(7806
keV) reaction from the nonlocal adiabatic potential analyses using different
NN potential models. The nonlocal model spectroscopic factors were deduced
so as to reproduce the local adiabatic model calculations of Ref. \cite{Mar15},
the solid black line in Fig. \ref{fig:8}.}
\centering
\begin{tabular}{p{2.4 cm}p{3.2 cm}p{2.9 cm}p{2.9cm}p{2.9cm} p{2.2 cm}}
\hline\hline
& Work of Ref. \cite{Mar15} &  \multicolumn{4}{c}{Present work} \\
\cline{2-6}
& & Hulth\'en & AV18 &CD-Bonn &RSC \\
\hline
$S_{l=0}$ & 9.3(19)$\times 10^{-3}$ & 8.6(1)$\times 10^{-3}$ & 8.2(2)$
  \times 10^{-3}$& 9.0(3)$\times 10^{-3}$ & 8.2(2)$\times 10^{-3}$  \\
$S_{l=2}$ & 6.8(14)$\times 10^{-2}$ & 5.8(2)$\times 10^{-2}$ & 3.3(2)$
  \times 10^{-2}$& 4.5(4)$\times 10^{-2}$ & 2.8(2)$\times 10^{-2}$  \\
$S_{l=0}/S_{l=2}$ & 0.14 & 0.15(1) & 0.25(2) & 0.20(2) & 0.29(2) \\
\hline\hline
\end{tabular}
\label{tab:sf}
\end{table*}

\subsection{Uncertainty of spectroscopic information extracted from
(d,p) reactions: $^{26}$Al$(d,p)^{27}$Al(7806 keV) case.}
\begin{figure}
\bigskip
\includegraphics[scale=0.35]{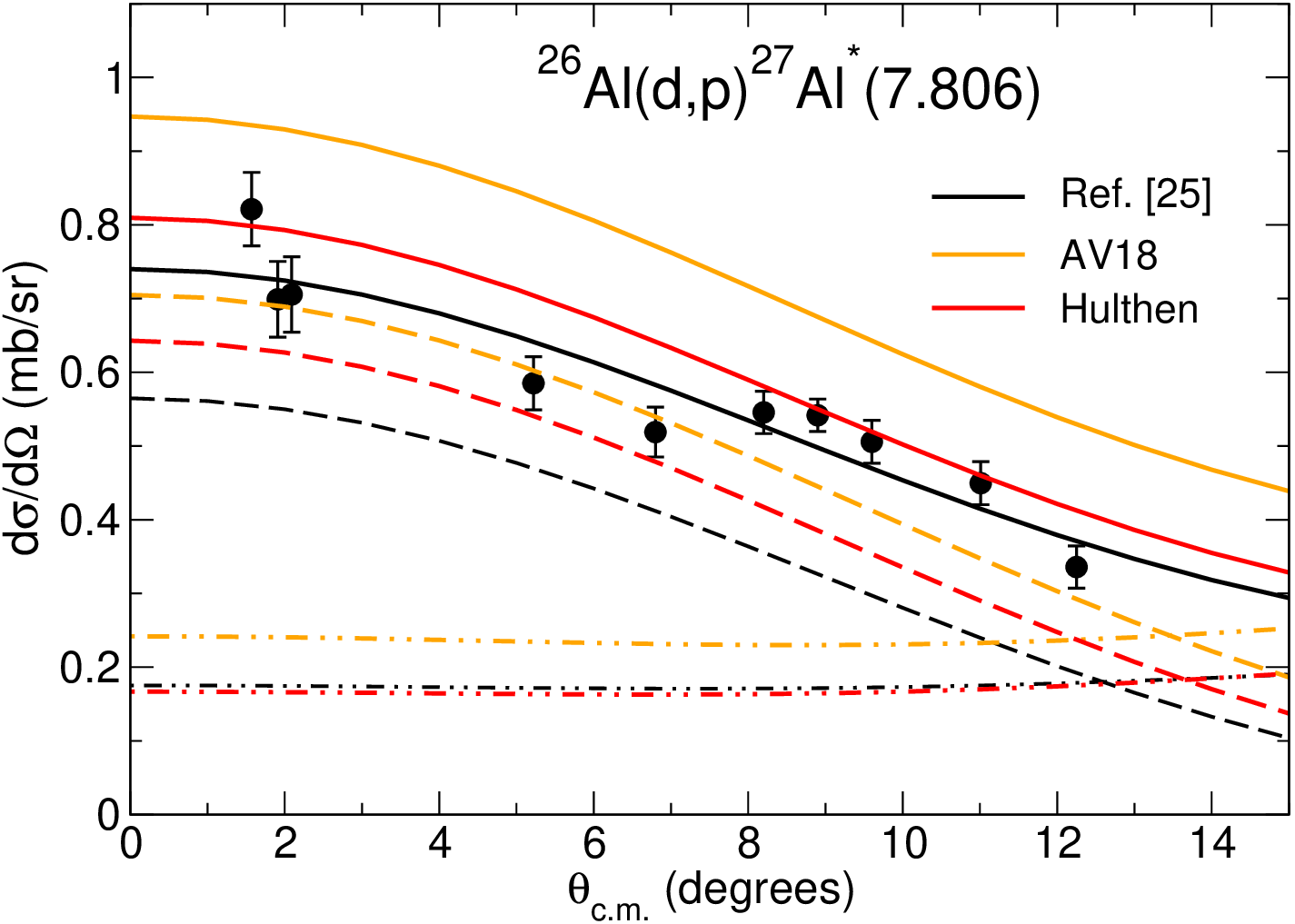}
\caption{\label{fig:8} Calculated $l=0$ (dashed lines) and $l=2$ (dot-dashed lines)
differential cross sections and their sums (solid lines) for the $^{26}$Al$(d,p)^{27}
$Al (7806 keV) reaction at 12 MeV. The red and orange curves result from nonlocal
potential analyses using the Hulth\'en and AV18 NN wave function models. These
calculations have been scaled by the spectroscopic factors deduced from the data
using the local potential analysis of Ref. \cite{Mar15}, shown by the black lines.}
\end{figure}

To illustrate how including the $D$-state in the nonlocal adiabatic model
can affect the spectroscopic factors extracted from $(d,p)$ reactions, we
present calculations of the $^{26}$Al$(d,p)^{27}$Al$^*$(7806 keV) reaction.
The reaction populates the mirror of an astrophysically important state in
$^{27}$Si, relevant to the destruction of $^{26}$Al in Wolf-Rayet and
Asymptotic Giant Branch stars \cite{Pai15,Mar15}. The differential cross
section for this transition is an (incoherent) combination of $l=0$ and
$l=2$ single-particle transfers; however, only the $l=0$ part is important
for characterizing the low-energy $^{26}$Al + $p$ resonance in the $^{27}$Si
mirror. Spectroscopic factors of $S_{l=0}$=9.3(19)$\times 10^{-3}$
and $S_{l=2}$=6.8(14)$\times 10^{-2}$ were deduced in Ref. \cite{Mar15}
from new high-precision data using an analysis that used the ADWA and
local global nucleon optical potentials \cite{KD02}. These calculated
cross sections (black curves) and the experimental data are shown in Fig.
\ref{fig:8}. The figure also shows the cross sections obtained from the
present nonlocal adiabatic model for two choices of NN potential, the
pure $S$-state Hulth\'en and the more realistic AV18 potential. All of
the calculated $l=0$ and $l=2$ transfer contributions have been scaled
using the spectroscopic factors of Ref. \cite{Mar15} above. The nonlocal
model calculations use the same geometries for the neutron bound states
potentials as in Ref. \cite{Mar15}, with radius parameters $r_0$ = 1.159
fm and 1.263 fm for the $l=0$ and $l=2$ states, respectively, diffuseness
$a_0=0.7$ fm and a spin-orbit potential depth $V_{SO}=6$ MeV.

The summed $l=0$ and $l=2$ partial cross sections are larger than the
experimental data and the earlier local-model calculations and hence the
deduced spectroscopic factors from the nonlocal model are smaller. The
revised spectroscopic factors from the present nonlocal analyses, fitted so
as to reproduce the angular distribution from the local analysis (the black
solid curve) are shown in Table \ref{tab:sf}. The errors shown for the
nonlocal calculations are those associated with this fit of the different
theoretical calculations and do not include the uncertainties associated
with the fitting to the data points, shown for the local-model analysis.
The spectroscopic factors obtained from analyses using the CD-Bonn \cite{CDB}
and Reid Soft Core (RSC) \cite{RSC} NN potentials are also tabulated. It
was shown in Ref. \cite{Bai16} that these CDB and RSC cross sections
essentially provide upper and lower bounds to those calculated with the
other NN potentials studied there. One notes that the revised $S_{l=0}$
are reduced by up to 12$\%$ from those of the local potential analysis
\cite{Mar15}. Moreover, while the $S_{l=2}$ obtained with the $S$-state
Hulth\'en potential are similar, the reduction can be greater depending
on the NN potential choice. Overall, we find this choice introduces an
$\approx$60$\%$ uncertainty in the deduced $S_{l=2}$, which is significantly
larger than the quoted experimental uncertainties. While important for
considerations of the structure of $^{27}$Al, this uncertainty does not
affect the main conclusion of Ref. \cite{Mar15} - that the $^{26}$Al($p,
\gamma)^{27}$Si capture and $^{26}$Al destruction mechanism in novae is
dominated by the $l=0$ channel.

\section{Conclusions}
We have extended the nonlocal adiabatic model of $A(d,p)B$ reactions to include the
deuteron $D$-state. Whereas adiabatic model deuteron channel potentials generated
from local nucleon optical potentials are insensitive to the deuteron $D$-state,
the nonlocality of the nucleon optical potentials emphasizes the high-momentum
parts of the deuteron wave functions in which the $D$-state component plays an
important role. As a result, the $(d,p)$ cross sections calculated in the nonlocal
adiabatic model are significantly affected by the $D$-state component.

We have presented exact calculations of the nonlocal adiabatic model wave functions.
To clarify the $D$-state dependence we have also performed lowest-order calculations
in which the nonlocal nucleon optical formfactors $U_{NA}$ are evaluated at the $n-p$
center of mass position. This approximation is shown to generate the leading modifications
to the $(d,p)$ cross sections and to provide insight into the physical picture. Namely,
it clarifies that the deuteron channel adiabatic potential can also be generated from
local nucleon optical potentials if these are evaluated at energies that are shifted
with respect to the usually-used value, $E_d/2$. Inclusion of the deuteron $D$-state,
through the use of realistic NN forces and deuteron wave functions, is shown to increase
this energy shift leading to shallower and less absorptive deuteron channel distorting
potentials compared to those calculated using a purely $S$-state deuteron wave function.
Cross sections calculated using this leading order approximation differ from the
exact calculations by 12$\%$ and 10$\%$ for deuteron wave function models with and
without a $D$-state, respectively.

The degree to which the significant $D$-state effects upon the central terms of the
deuteron channel interaction affect the $(d,p)$ cross section magnitudes and angular
distributions depend on the transferred angular momentum and the neutron separation
energy of the final state. Our calculations show that they can be as large as 50$\%$
for some cases and, when two values of $lj$ are allowed by the selection rules, they
add ambiguity to the interpretation of experimental data. For example, when the
$^{26}$Al$(d,p)^{27}$Al reaction populating the astrophysically-relevant $^{27}$Al
(7806 keV) state is analyzed in our nonlocal model, the deduced spectroscopic factors
for both $l=0$ and $l=2$ transfers are reduced. While an analysis without the deuteron
$D$-state reduces these spectroscopic factors by no more than 16$\%$, inclusion of
the $D$-state results in a dramatic reduction of the extracted $l=2$ spectroscopic
strength, by up to a factor of two. The uncertainty associated with different $S$ +
$D$-state deuteron ground state wave function models is of order 60$\%$. By contrast,
the effects of spin-tensor potential terms induced by the deuteron $D$-state, even when
the nucleon optical potentials are central, are included in the calculated deuteron
channel distorted waves and are shown to have negligible effects on the calculated
$(d,p)$ cross sections.

Apart from the $^{26}$Al$(d,p)^{27}$Al reaction case, we have presented $D$-state results
only for the AV18 wave function. The other NN models studied predict smaller $D$-state
effects \cite{Bai16}, so the present results can be considered as a reasonable upper
limit. Calculations for $^{26}$Al using the CD-Bonn potential give the smallest
$(d,p)$ cross sections of the models studied and could similarly be considered as a
lower limit, but nevertheless show the importance of the $D$-state contribution to the
spectroscopic factors obtained.

Finally, the present study has used energy-independent nonlocal nucleon potentials.
Explicit energy-dependence of the nonlocal nucleon optical potentials, discussed in
Ref. \cite{Joh14}, may significantly modify model predictions - as was shown in
Ref. \cite{Wal16} for the case of a purely $S$-state deuteron.

\appendix*
\section{Multipole expansions}
The multipole expansion of the nonlocal deuteron channel potential ${\cal U}_{dA}$ in
Eq. (\ref{Udafinal}), expressed as a function of the variables $\ve{R}$ and $\ve{S}$,
includes the function $v^a_{ l_1 l_2} (R,S)$ of the radial variables. This is given
by the following expression
\beq
&&v^a_{ l_1 l_2} (R, S) =  8 H(2S) \hat{l}_1 \hat{J}_d\sum_{l_d l'_d k } (l_1 0 k0|
l'_d0) \eol &&\  \times\, \hat{l}_d' \hat{k}\, W(a l'_d l_2 k;l_d l_1) W(a l_d J_d
s_d;l'_d J_d) \eol &&\ \times \int_0^\infty dx \, x\, \tilde{\upsilon}_{k l_2}^{(l_d)}
(x,2S) \tilde{U}_{l_1}({x}/{2},R) u_{l'_d}(x) .
\eeqn{A1}
Inspection shows that the angular momentum couplings in the second Racah coefficient,
with $J_d=s_d=1$ and $l_d, l'_d=0,2$, restrict the spin tensor terms in ${\cal U}_{dA}$
of Eq. (\ref{Udafinal}) to $a=0$ (central) and $a=2$ (rank-2 tensor) components.

In Eq. (\ref{A1}), $\tilde{\upsilon}_{k_1 k_2}^{(l_d)}$ arises from the multipole
expansion of the radial components, $v_{l_d}$, of the deuteron vertex function, $V_{np}
\phi_0$, as detailed in and following Eq. (\ref{dwf}). Explicitly,
\beq
&&\tilde{\upsilon}_{k_1 k_2}^{(l_d)}(x,2S) = \frac{(-1)^{k_2}}{2\la \phi_0|V_{np}|\phi_0
\ra} \sum_{e,{c+d=l_d}} \hat{e}^2 \left[\frac{(2l_d + 1)!} {(2c)!(2d)!}\right]^{\frac{1}
{2}} \eol &&\ \times\, (c0e0|k_10) (d0e0|k_20) W(k_1 e l_d d;c k_2) \eol &&\ \times\,
x^{c} (2S)^{d} \int^1_{-1}\frac{v_{l_d}\left(\left|\ve{x}-2\ve{S}\right|\right)} {\left|
\ve{x}-2\ve{S} \right|^{l_d+1}} \, P_e(\mu)\, d\mu
\eeq
with $\mu = \ve{x}\cdot \ve{S}/xS$.

Finally, the $\tilde{U}_l$ in Eq. (\ref{A1}) are the multipoles
of the sum the proton and neutron nonlocal potential form factors, i.e. $\tilde{U}_l
({x}/{2},R)=\tilde{U}_l^n({x}/{2},R)+\tilde{U}_l^p ({x}/{2},R)$, where
\beq
\tilde{U}_l^N({x}/{2},R) = \int_{-1}^1 d\mu \,U_{NA}\big(\big|\ve{x}/2-
\ve{R}\big|\big)\,  P_l(\mu) \ \
\eeq
with $\mu = \ve{x}\cdot \ve{R}/xR$.

\section*{Acknowledgement}
We are grateful to Professor R.C. Johnson for many useful discussions. This
work was supported by the United Kingdom Science and Technology Facilities
Council (STFC) under Grant No. ST/L005743/1.


\begin{thebibliography}{50}
\bibitem{Austern} N. Austern, Direct Nuclear Reaction Theories, Wiley, New York, 1970.
\bibitem{JT} R.C. Johnson and P.C. Tandy, Nucl. Phys. {\bf A235}, 56 (1974).
\bibitem{Fesh} H. Feshbach, Ann. Rev. Nucl. Sci. {\bf 8}, 49 (1958).
\bibitem{PBuck} F. Perey and B. Buck, Nucl. Phys. {\bf 32}, 353 (1962).
\bibitem{Per63} F. Perey, Direct interactions and nuclear reaction mechanisms
(Gordon and Breach, N.Y. 1963), p.125
\bibitem{Tit14} L.J.Titus, F.M.Nunes, Phys.Rev. C {\bf 89}, 034609 (2014).
\bibitem{Ros15} A. Ross, L.J. Titus, F.M. Nunes, M.H. Mahzoon, W.H. Dickhoff,
R.J. Charity, Phys.Rev. {\bf 92}, 044607 (2015).
\bibitem{Tim13a} N.K. Timofeyuk and R.C. Johnson, Phys. Rev. Lett. {\bf 110},
112501 (2013).
\bibitem{Tim13b} N.K. Timofeyuk and R.C. Johnson, Phys. Rev. C. {\bf 87},
064610 (2013).
\bibitem{Tit16} L.J.Titus, F.M.Nunes, G.Potel, Phys. Rev. C {\bf 93}, 014604 (2016).
\bibitem{Ros16} A. Ross, L.J. Titus, F.M. Nunes, Phys. Rev. C {\bf 94}, 014607 (2016).
\bibitem{Del09} A. Deltuva, Phys.Rev. C {\bf 79}, 021602 (2009).
\bibitem{Hul} L. Hulth\'en and M. Sugawara, in Handbuch der Physik, Ed. by S. Flugge
(Springer-Verlag, Berlin, 1957), p. 1.
\bibitem{Joh67} R.C. Johnson, Nucl. Phys. {\bf A90}, 289 (1967).
\bibitem{Del74} G. Delic and B.A. Robson, Nucl. Phys. {\bf A232}, 493 (1974).
\bibitem{Bai16} G.W.Bailey, N.K. Timofeyuk and J.A. Tostevin, Phys. Rev. Lett.
{\bf 117}, 162502 (2016).
\bibitem{Pan13} D.Y. Pang, N.K. Timofeyuk, R.C. Johnson and J.A. Tostevin, Phys. Rev.
C {\bf 87}, 064613 (2013).
\bibitem{JS}  R.C. Johnson and P.J.R. Soper, Phys. Rev. C {\bf 1}, 976 (1970).
\bibitem{AV18} R.B. Wiringa, V.G.J. Stoks and R. Schiavilla, Phys. Rev. C. {\bf 51},
38 (1995).
\bibitem{CH89} R.L. Varner et al, Phys. Rep. {\bf 201}, 57 (1991).
\bibitem{Watanabe} S. Watanabe, Nucl. Phys. {\bf 8}, 484 (1958).
\bibitem{Ngu10} N. B. Nguyen, F. M. Nunes, and R. C. Johnson, Phys. Rev. C {\bf 82},
014611 (2010).
\bibitem{twofnr} J.A. Tostevin, University of Surrey version of the code {\sc twofnr}
(of M. Toyama, M. Igarashi and N. Kishida) and code {\sc front} (private communication).
\bibitem{GR76} M.M. Giannini and G. Ricco, Ann. Phys. {\bf 102}, 458 (1976).
\bibitem{Mar15} V. Margerin, G. Lotay, P. J. Woods, M. Aliotta, G. Christian, B. Davids,
T. Davinson, D.T. Doherty, J. Fallis,  D. Howell, O.S. Kirsebom, D.J. Mountford,
A. Rojas, C. Ruiz,  and J.A. Tostevin, Phys. Rev. Lett. {\bf 115}, 062701 (2015).
\bibitem{Pai15} S. D. Pain, D. W. Bardayan, J. C. Blackmon, S. M. Brown, K. Y. Chae,
K. A. Chipps, J. A. Cizewski, K. L. Jones, R. L. Kozub, J. F. Liang, C. Matei, M. Matos,
B. H. Moazen, C. D. Nesaraja, J. Okolowicz, P. D. O'Malley, W. A. Peters, S. T. Pittman,
M. Ploszajczak, K. T. Schmitt, J. F. Shriner, Jr., D. Shapira, M. S. Smith, D. W. Stracener,
and G. L. Wilson, Phys. Rev. Lett. {\bf 114}, 212501 (2015).
\bibitem{KD02} A.J. Koning, J.P. Delaroche,  Nucl.Phys. {\bf A713}, 231 (2003).
\bibitem{CDB} R. Machleidt, Phys. Rev. C. {\bf 63} 024001 (2001).
\bibitem{RSC} R.V. Reid, Ann. Phys. {\bf 50}, 411 (1968).
\bibitem{Joh14} R.C. Johnson and N.K. Timofeyuk, Phys. Rev. C {\bf 89}, 024605 (2014).
\bibitem{Wal16} S.J. Waldecker and N.K. Timofeyuk, Phys. Rev. C {\bf 94}, 034609 (2016).
\end{thebibliography}
\end{document}